\documentclass[twocolumn,twoside,slac_two]{revtex4}

\usepackage{graphicx}
\usepackage{fancyhdr}
\def\kr{\hbox{ \raisebox{-1.0mm}{$\stackrel{<}{\sim}$} }}

\begin{document}


\title{Evidence for Supernova light in all Gamma-Ray Burst afterglows}

\author{A. Zeh, S. Klose}
\affiliation{Th\"uringer Landessternwarte Tautenburg, 07778 Tautenburg,
  Germany}

\author{D. H. Hartmann}
\affiliation{Department of Physics and Astronomy, Clemson University,
  Clemson, SC 29634-0978}

\begin{abstract}
We present an update of our systematic analysis of all Gamma-Ray Burst (GRB)
afterglow data, now published through the end of 2004, in an attempt to detect
the predicted supernova light component. We fit the observed photometric light
curves as the sum of an afterglow, an underlying host galaxy, and a supernova
component. The latter is modeled using published $UBVRI$ light curves of SN
1998bw as a template. The total sample of afterglows with established
redshifts contains now 29 bursts (GRB 970228 - GRB 041006). For 13 of them a
weak supernova excess (scaled to SN 1998bw) was found.  In agreement with our
earlier result \cite{Zeh2004} we find that also in the updated sample all
bursts with redshift \kr 0.7 show a supernova excess in their afterglow light
curves. The general lack of a detection of a supernova component at larger
redshifts can be explained with selection effects. These results strongly
support our previous conclusion based on all afterglow data of the years 1997
to 2002 \cite{Zeh2004} that in fact \it all \rm afterglows of long-duration
GRBs contain light from an associated supernova.
\end{abstract}

\maketitle

\thispagestyle{fancy}

\section{Introduction}

Significant progress towards understanding the nature of GRBs and their
progenitors came with the discovery of GRB afterglows in 1997 \cite{Groot1997,
vanParadies1997}. First observational evidence for the underlying source
population was provided by GRB 970828, which showed a bright X-ray afterglow
but no optical counterpart down to faint magnitudes \cite{Groot1998}. This led
to the suggestion that the optical light was blocked by cosmic dust in the GRB
host galaxy, linking the burster to a dusty  star-forming region, i.e., most
likely to the explosion of a massive star \cite{Paczynski1998}. The discovery
of a near-by type Ibc supernova (SN 1998bw) in the error circle of the X-ray
afterglow for GRB 980425 \cite{Galama1998, Kulkarni1998}, provided
strong support for this idea, and is consistent with our current understanding
of type Ibc SNe and their progenitors (e.g., \cite{Fryer1999, Heger2003}).

From the observational site, the supernova picture is further supported by the
fact that all GRB hosts are star-forming, and in some cases even star-bursting
galaxies (e.g., \cite{Frail2002, Sokolov2001}). Evidence for host
extinction by cosmic dust in GRB afterglows
and the discovery of an ensemble  of optically 'dark
bursts' (for a recent discussion, see \cite{Fynbo2001, Klose2003,
Lazzati2002}) also is consistent with the picture that GRB progenitors
are young, massive stars. Furthermore, for several GRB afterglows X-ray lines
may hint at a period of nucleosynthesis preceding or accompanying the burst
\cite{Antonelli2000, Lazzati1999, Meszaros2001}. The
positions of the afterglows with respect to their hosts also favors a relation
to young, massive stars to GRBs \cite{Bloom2002}.

As a natural consequence of a physical relation between the explosion of
massive stars and GRBs supernova light should contribute to the afterglow
flux, and even  dominate under favorable conditions. The most convincing
example is GRB 030329 \cite{Peterson2003} at $z$=0.1685 \cite{Greiner2003a}
with spectral confirmation of supernova light in its afterglow
\cite{Hjorth2003, Kawabata2003, Matheson2003,
Stanek2003}. Spectroscopic evidence for SN light in a GRB afterglow was
later also reported for  GRB 021211 \cite{dellaValle2003}, GRB 031203
\cite{Malesani2004} and most recently for XRF 020903 \cite{Soderberg2005}.

\begin{figure}
\includegraphics[width=85mm]{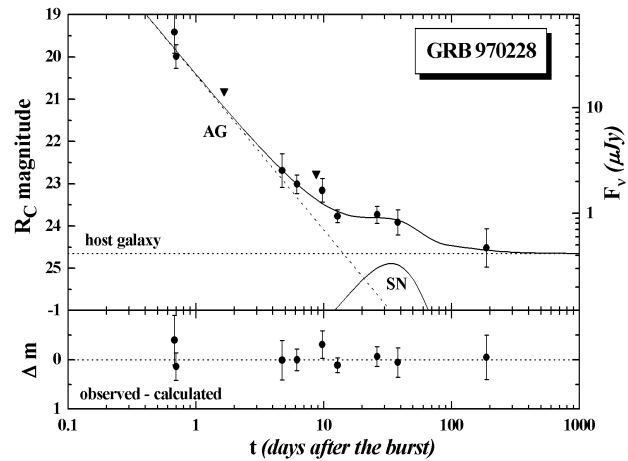}
\caption{
The late-time bump in the optical afterglow of GRB 970228, the first optical
afterglow ever found (data collected from the literature). Interpreted as a
signature from a SN explosion makes this event one of the most distant
core-collapse SNe ever seen at that time. See also \cite{Reichart1999,
Galama2000}.}
\label{970228}
\end{figure}

In contrast to direct spectroscopic evidence, several cases of \it photometric
\rm indication of extra light in GRB afterglows have been reported, starting
with the pioneering work on GRB 980326 \cite{Bloom1999}. Inspired by this
finding, the  discovery of extra light in archived data of the afterglow of
GRB 970228 \cite{Reichart1999, Galama2000} made it clear that a search
for late-time bumps in optical afterglow light curves provides a powerful tool
to constrain or even reveal the nature of the underlying sources. Since then
various groups successfully fit SN 1998bw templates to explain these late-time
bumps (e.g., \cite{Dado2002}), the most convincing case being that of GRB
011121 \cite{Bloom2002, Garnavich2003, Greiner2003b}.

The goal of our study is to search for supernova bumps in GRB afterglow light
curves using a systematic approach, allowing us to draw statistically founded
conclusions on the physical properties of this new class of GRB-SNe in
particular and on the GRB progenitors in general. We collected from the
literature all available photometric data on GRB afterglows (including our own
data), checked them for photometric consistency, and re-analyzed the data in a
consistent manner. Here we report on the status of our study for all bursts
that occurred by the end of 2003, supplementing and expanding our previous
results (\cite{Zeh2004}, in the  following paper I).

\section{Numerical approach \label{numerics}}

We model the light curve of the optical transient (OT) following a GRB as a
composite of afterglow (AG) light, supernova (SN) light, and constant light
from the underlying host galaxy.  The flux density, $F_\nu$, at a frequency
$\nu$ is then given by
\begin{equation}
 F_\nu^{\rm OT}(t) = F_\nu^{\rm AG}(t) + k \, F_\nu^{\rm SN}(t/s)
 + F_\nu^{\rm host}\,.
\label{ot}
\end{equation}
Here, the parameter $k$ describes the observed brightness ratio (in the host
frame, i.e., including the cosmological $K$-correction)
between the GRB-supernova, and the SN template (SN 1998bw) in the
considered photometric band (in the observer frame). We allowed $k$ to be
different in every photometric band, but within a band independent of
frequency. The parameter $s$ is a stretch factor with respect to the used
template. We have also explored  the consequences of a shift in time between
the onset of the burst and the onset of the supernova explosion, as implied by
some theoretical models \cite{Vietri2000}. Then, in Eq.~(\ref{ot}) $F_\nu^{\rm
SN}(t/s)$ was replaced by $F_\nu^{\rm SN}(t+\tau)$.  Here, $\tau=0$ refers to
GRB 980425/SN 1998bw \cite{Iwamoto1998}. If  $\tau<0$ the SN preceded the
onset of the GRB.

Following \cite{Beuermann1999} and \cite{Rhoads2001}, we describe the
afterglow light curve by a broken power-law,
\begin{equation}
   F_\nu^{\rm AG}(t) = \mbox{const}\
   [(t/t_b)^{\alpha_1\,n}+(t/t_b)^{\alpha_2\,n}]^{-1/n}\,,
\label{AG}
\end{equation}
with const=$2^{1/n} \, F_\nu^{\rm AG}(t_b).$ Here $t$ is the time after the
burst (in the observer frame),  $\alpha_1$  is the pre-break decay slope of
the afterglow light curve, $\alpha_2$ is the post-break decay slope, and $t_b$
is the break time. The parameter $n$ characterizes the sharpness of the break;
a larger $n$ implies a sharper break.  If no break is seen in the data then
$\alpha_1 = \alpha_2$ and  Eq.~(\ref{AG}) simplifies correspondingly.

The results of this numerical procedure were compared with corresponding
results published by others \cite{Dado2002, Bloom2002}, and we found close
agreement. We used this procedure to \it predict \rm the color evolution of
GRB 030329/SN 2002dh \cite{Zeh2003}, and obtained a very good numerical fit
for the light curves of GRB-SN 011121 \cite{Greiner2003b}. The limitations of
the procedure are given by the chosen photometric band in combination with the
redshift of the burster. Once we can no longer interpolate in between the
$UBVRI$ bands, but have to extrapolate into the UV domain
(cf. \cite{Bloom1999}), results become less accurate. For more details see
paper I.

\begin{figure}
\includegraphics[width=85mm]{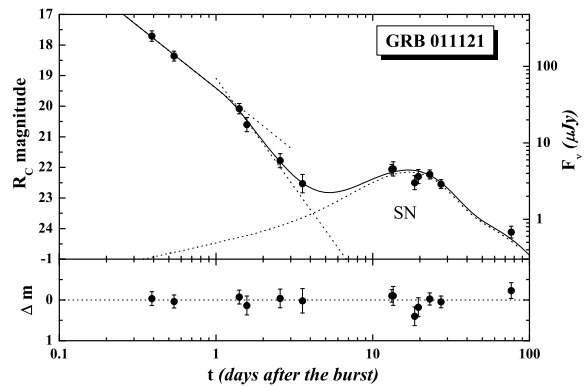}
\caption{
The afterglow of GRB 011121 showed a very clear signature of a late-time bump
rising some days after the burst. Shown here are data obtained with the
telescopes at ESO, Chile, and with the Hubble Space Telescope
\cite{Bloom2002}). The bump can be  modeled well by an underlying SN component
at the redshift of the burster ($z$=0.36) with a peak luminosity of about 80\%
of the peak luminosity  of SN 1998bw. Note that in the figure the  flux from
the underlying host  galaxy was subtracted from the data (for details, see
\cite{Greiner2003b}).}
\label{011121}
\end{figure}

\begin{figure}
\includegraphics[width=85mm]{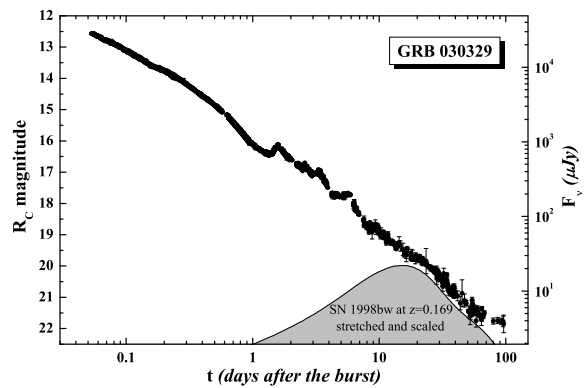}
\caption{Sketch of the
hidden SN bump in the afterglow of GRB 030329. Various re-brightening
episodes of the genuine afterglow, in combination with a relatively late
break-time of the  light curve, made the photometric signature for the
underlying SN explosion very small. Presumably, with no spectroscopic evidence
at hand, the SN had easily been missed in the data. For the light curve fit
$\alpha_2$ was fixed at 2.5.}
\end{figure}

Before performing a numerical fit, the observational data was corrected for
Galactic extinction along the line of sight using the COBE maps
\cite{Schlegel1998}. This also holds for SN 1998bw, where we assumed $E(B-V)$=
0.06 mag. We calculated the Galactic visual extinction according to $A_V^{\rm
Gal}$ = 3.1 $E(B-V)$, whereas the extinction in $U$ and $B$ were obtained via
\cite{Rieke1985}, and in $R_c$ and $I_c$  by means of the numerical functions
compiled in \cite{Reichart2001}.

\begin{table}
\caption{The input sample of GRB afterglows. Redshifts were taken from
the literature. \label{allgrbs}}
\vspace{0.3cm}
\begin{tabular}{|ll|ll|ll|ll|}
\hline \noalign{\smallskip}
GRB & $z$ & GRB & $z$ & GRB & $z$ & GRB & $z$ \\
\noalign{\smallskip} \hline
\noalign{\smallskip}
970228 & 0.695 & 991216 & 1.02  & 011121 & 0.362 & 030226 & 1.986 \\
970508 & 0.835 & 000301C& 2.04  & 011211 & 2.140 & 030323 & 3.372 \\
971214 & 3.42  & 000418 & 1.118 & 020405 & 0.69  & 030328 & 1.520 \\
980703 & 0.966 & 000911 & 1.058 & 020813 & 1.25  & 030329 & 0.169 \\
990123 & 1.600 & 000926 & 2.066 & 020903 & 0.251 & 030429 & 2.658 \\
990510 & 1.619 & 010222 & 1.477 & 021004 & 2.3   & 031203 & 0.106 \\
990712 & 0.434 & 010921 & 0.450 & 021211 & 1.01  & 041006 & 0.716 \\
991208 & 0.706 &        &       &        &       &        &       \\
\noalign{\smallskip} \hline
\end{tabular}
\end{table}

Most of the light curves we investigated have been followed in more than one
photometric band. For each of these GRBs we chose the best-sampled light curve
as a reference light curve for the fit in the other photometric bands. In all
cases this was the $R$ band light curve.  We always assumed that \it
afterglows \rm are achromatic, in reasonable agreement with observational data
(e.g., \cite{Harrison1999, Klose2004}).
For every individual GRB, the afterglow parameters
$\alpha_1, \alpha_2, t_b$, and $n$ (Eq.~\ref{AG}) are then the same for all
photometric bands. Consequently, once we fit the reference light curve of an
optical transient and deduced the corresponding afterglow parameters, we
treated them as fixed parameters when fitting the  light curves of the optical
transient in other photometric bands. In the fit  the degrees of freedom are
reduced correspondingly.

\section{Results and Discussion \label{results}}

The input sample consists of 29 bursts (GRB 970228 - GRB 041006) with
established  redshifts and good enough photometric data in order to search for
a late-time bump in their afterglows (Table~\ref{allgrbs}), with the most
recent data for GRB 041006 \cite{Stanek2005}. These are eight bursts more
than in our previous study for all bursts observed by the end of 2002 (paper
I). Among these 29 bursts are 13 for which a late-time bump was found. This
includes now also XRF 020903 (as already noted in \cite{Soderberg2002} and now
spectroscopically confirmed \cite{Soderberg2005}). Note that the requirement
of a known redshift excludes GRB 980326 as well as XRF 030723 from this list,
which both showed a strong late-time bump. On the other hand, as already noted
in \cite{Bloom1999}, one can in principle constrain the redshift of a burster
by fitting a redshifted SN component to the observed late-time bump in its
afterglow light curve.

\begin{figure}[h!]
\hspace*{-0.5cm}
\includegraphics[width=90mm]{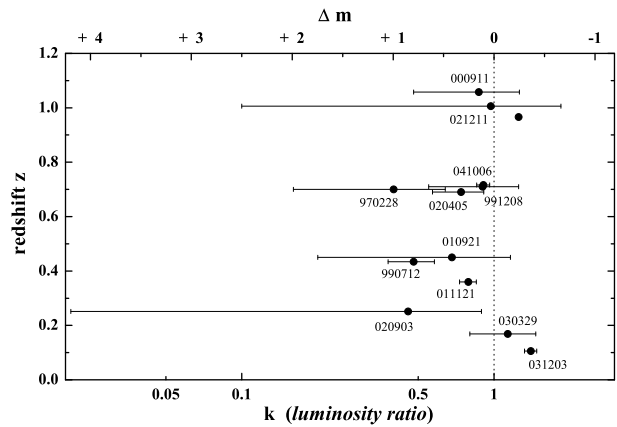}
\caption{
The deduced peak luminosities of all GRB-SNe in units of the peak  luminosity
of SN 1998bw. All data refer to the $R_c$ band (in the observer frame). The
dotted line corresponds to SN1998bw. The parameter $\Delta m$ equals  --2.5
log $k$, which  measures the magnitude difference at maximum light between the
GRB-SN and SN 1998bw in the corresponding wavelength regime. Note that the
data are not corrected for a possible extinction in the GRB host galaxies. GRB
980703 is not included here and in the following figures because in this case
the stretch factor was not allowed to vary freely. \label{k}}
\end{figure}

\begin{figure}
\hspace*{-0.5cm}
\includegraphics[width=90mm]{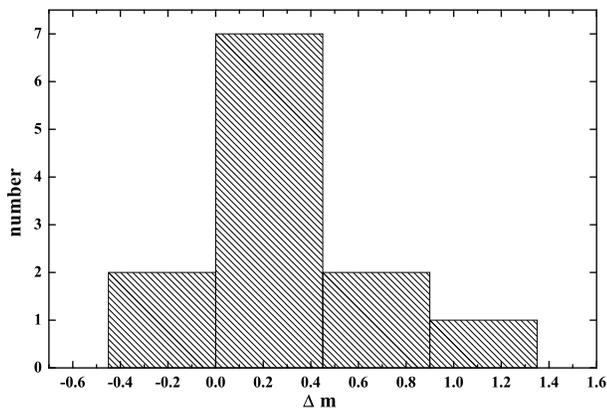}
\caption{
The distribution of the peak brightness of all GRB-SNe in units of the
corresponding peak brightness of  SN 1998bw in the $R_c$ band (in the observer
frame).  The corresponding stretch factor is shown in Fig.~\ref{s}.
\label{brightness}}
\end{figure}

Again, our key finding is photometric evidence of a late-time bump \it in all
\rm GRB afterglows with a redshift $z\kr$0.7.  We interpret this bump as
light from an underlying supernova, and model this component as a redshifted
version of SN 1998bw. The deduced luminosities for these GRB-SNe   (not
including extinction corrections for the host galaxy),
normalized to  SN 1998bw are listed in
Table~\ref{res}.  The width of the distribution of the SN peak luminosities
(in units of the  peak luminosity of  SN 1998bw) spans over 2 photometric
magnitudes with a pronounced maximum around $k=0.5...1$ (Figs.~\ref{k},
\ref{brightness}). The potential SN related to the X-ray flash 020903 is not
unusual with respect to its peak luminosity. Interestingly, SN 1998bw is at
the bright end of the GRB-SNe distribution (as already noted in paper I). Only
the SNe related to GRBs 030329 and 031203 might have been slightly more
luminous at peak brightness.  No correlation was found of the deduced SN
luminosities with the redshift or any afterglow parameter. Note, however, that
we cannot exclude the existence of such a correlation since in most cases when
a SN was found there is a lack of early time data in the optical light curve
(resulting in an unknown break time $t_b$ and, hence, an unknown parameter
$\alpha_1$; Eq.~\ref{AG}).

\begin{figure}
\hspace*{-0.5cm} \includegraphics[width=90mm]{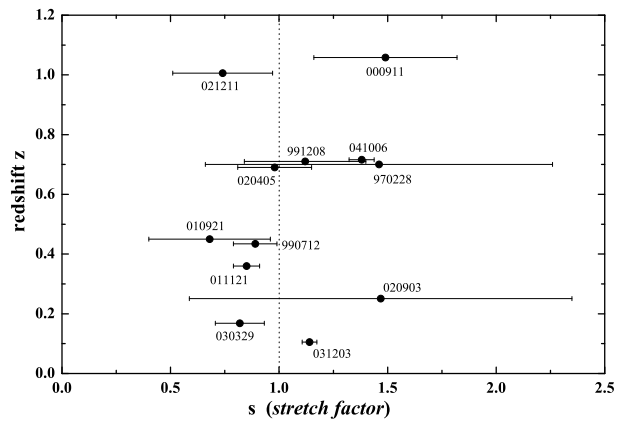}
\caption{
The distribution of the corresponding parameter $s$ (Eq.~\ref{ot}) describing
a stretching of the SN light curve relative to those of SN 1998bw (for which
by definition $s=1$, dotted line). In general, $s<0$ ($s>0$) means that the
evolution of the light curve of the SN was slower  (faster) than those of the
light curve of SN 1998bw in the corresponding wavelength band. The mean value
is $s=1.0$. \label{s}}
\end{figure}

\begin{table*}
\caption{
Best-fit parameters for the SN component found in GRB afterglows with known
redshift.  Columns: (1) and (2) GRB and redshift; (3) photometric band, in
which the light curve was fitted; (4) central wavelength of the photometric
band in the host frame in units of nm, adopting a wavelength of 659 nm for the
$R_c$ band; (5) peak luminosity of the fitted SN component in the
corresponding wavelength band (observer frame) in units of SN 1998bw, after
correction for Galactic extinction; (6) stretch factor $s$ (Eq.~\ref{ot}); (7)
goodness of fit per degree of freedom; (8) and (9) the same as (5) and (7) for
$s=1$.  The low $\chi^2$/d.o.f. for GRB 970228 is due to the small number of
data points.
Note that we reduced all data with our own numerical procedure, so that slight
differences to the results obtained by others do naturally exist.
\label{res}}
\vspace*{3mm}
\renewcommand{\tabcolsep}{8pt}
\begin{tabular}{lcccccccc}
\hline \hline \\[-2mm]
\textbf{GRB} & \textbf{$z$} & \textbf{band} & \textbf{$\lambda_{\rm host}$}&
\textbf{$k$} & \textbf{$s$} & \textbf{$\chi^2_{\rm d.o.f.}$} &
\textbf{$k$ if $s$=1} & \textbf{$\chi^2_{\rm d.o.f.}$}\\[3mm]\hline
970228 & 0.695 & $R_c$ & 389 & 0.40$\pm$0.24 & 1.46$\pm$0.80 & 0.70 & 0.33$\pm$0.30 & 0.71\\
980703 & 0.966 & $R_c$ & 335 & --            & --            & --   & 1.66$\pm$1.22 & 0.78\\
990712 & 0.434 & $R_c$ & 459 & 0.48$\pm$0.10 & 0.89$\pm$0.10 & 1.00 & 0.43$\pm$0.08 & 1.01\\
991208 & 0.706 & $R_c$ & 386 & 0.90$\pm$0.35 & 1.12$\pm$0.28 & 1.64 & 1.02$\pm$0.32 & 1.56\\
000911 & 1.058 & $R_c$ & 320 & 0.87$\pm$0.39 & 1.49$\pm$0.33 & 0.75 & 0.51$\pm$0.43 & 1.14\\
010921 & 0.450 & $R_c$ & 454 & 0.68$\pm$0.48 & 0.68$\pm$0.28 & 0.42 & 0.43$\pm$0.10 & 0.78\\
011121 & 0.360 & $R_c$ & 484 & 0.79$\pm$0.06 & 0.85$\pm$0.06 & 0.92 & 0.74$\pm$0.05 & 1.32\\
020405 & 0.695 & $R_c$ & 389 & 0.74$\pm$0.17 & 0.98$\pm$0.17 & 5.26 & 0.72$\pm$0.11 & 4.86\\
020903 & 0.251 & $R_c$ & 527 & 0.46$\pm$0.44 & 1.47$\pm$0.88 & 1.52 & 0.37$\pm$0.41 & 1.22\\
021211 & 1.006 & $R_c$ & 328 & 0.97$\pm$0.87 & 0.74$\pm$0.23 & 2.68 & 0.52$\pm$0.34 & 2.65\\
030329 & 0.169 & $R_c$ & 563 & 1.13$\pm$0.33 & 0.82$\pm$0.13 & 3.10 & 0.98$\pm$0.01 & 4.49\\
031203 & 0.106 & $R_c$ & 596 & 1.65$\pm$0.41 & 1.14$\pm$0.16 & 0.04 & 1.75$\pm$0.19 & 0.24\\
041006 & 0.716 & $R_c$ & 384 & 0.91$\pm$0.05 & 1.38$\pm$0.06 & 1.27 & 1.21$\pm$0.07 & 1.95\\
\hline \hline\\[1mm]
\end{tabular}
\end{table*}

Figure~\ref{s} shows the distribution of the corresponding stretch  factor
$s$. Since no fit was possible for the afterglow of GRB 980703 with $s$ being
a free parameter, this burst is not included in Figs.~\ref{k}-\ref{s}. The
mean value of $s$ is 1.0, i.e., identical to SN 1998bw.

Instead of introducing a stretch factor to have more freedom in the variety of
GRB-SNe, one can also follow \cite{Vietri2000} and search for evidence of a
time delay between the burst and the SN. According to this model, GRBs are
the result of delayed black hole formation, which implies that the
core-collapse and its subsequent supernova may significantly precede the
burst. The delay could be of order months to years \cite{Vietri2000}, or
perhaps as short as hours \cite{Woosley2002}. For only two of the SN light
curves the fit indeed improved if we allowed for a shift in time between the
onset of the burst and the onset of the SN (GRBs 990712, 011121). The offsets
never exceeded 5 days, and were both negative and positive. However, the
uncertainties in this parameter are large, due to the poorly sampled shape of
the underlying supernova (e.g., \cite{Garnavich2003}).

\section{Summary and Conclusions}

Since the first clear evidence for extra light in a GRB afterglow light curve
(GRB 980326; \cite{Bloom1999}), there is growing evidence for several such
cases. Our key finding is photometric evidence of a late-time bump in \it all
\rm afterglows with a redshift $z\kr$0.7, including those of the year 2003
(GRBs 030329 and 031203) and year 2004 (GRB 041006;
\cite{Stanek2005}). For larger redshifts the
data is usually not of sufficient quality, or the SN is simply too faint, in
order to search for such a feature in the late-time afterglow light
curve. This extra light is modeled well by a supernova component, peaking
$(1+z)(15...20)$ days after a burst. This, together with the spectral
confirmation of SN light in the afterglows of GRB 021211, 030329, and 031203
further supports the view that in fact $all$ long-duration GRBs show SN bumps
in their late-time optical afterglows. Given the fact that a strong late-time
bump was also found for XRF 030723 \cite{Fynbo2004} and a less strong bump for
XRF 020903 (but with spectroscopic confirmation of
underlying SN light \cite{Soderberg2005}) might
indicate that this conclusion holds also for X-ray flashes
(even though the finding of XRF-SNe might be more
difficult; see \cite{Soderberg2005}).


\begin{acknowledgments}
S.K. and A.Z. acknowledge financial support by DFG grant KL 766/11-1.
D.H.H. acknowledges support for this project under NSF grant INT-0128882.
We thank N. Masetti and E. Palazzi, Bologna, for providing
host-subtracted data on the afterglow of GRB 020405. This work has
benefited from the GCN data base maintained by S. Barthelmy at NASA and the
\it GRB big table \rm maintained by J. Greiner, Max-Planck-Institut f\"ur
extraterrestrische Physik, Garching.
\end{acknowledgments}

\bigskip


\end{document}